\documentclass[aps,prb,twocolumn,preprintnumbers, showpacs,superscriptaddress]{revtex4-1}
\usepackage[final]{graphicx}
\usepackage{dcolumn}
\usepackage{multirow}
\usepackage{bm}
\usepackage{float}
\usepackage{amsmath}
\usepackage{xfrac}
\usepackage{amsfonts}
\usepackage{amssymb}
\usepackage{mathrsfs}
\usepackage[latin1]{inputenc}
\usepackage{siunitx}

\DeclareTextSymbol{\degre}{T1}{6}
\DeclareTextSymbol{\degre}{OT1}{23}

\begin{document}

\bibliographystyle{unsrt}

\title{Free electron properties of metals under ultrafast laser-induced electron-phonon nonequilibrium: a 
first-principles study}

\author{E. B\'{e}villon}
\affiliation{Laboratoire Hubert Curien, UMR CNRS 5516, Université de Lyon,\\ Université Jean-Monnet 42000,
Saint-Etienne, France}
\author{J.P. Colombier}
\email{jean.philippe.colombier@univ-st-etienne.fr}
\affiliation{Laboratoire Hubert Curien, UMR CNRS 5516, Université de Lyon,\\ Université Jean-Monnet 42000,
Saint-Etienne, France}
\author{V. Recoules}
\affiliation{CEA-DIF, 91297 Arpajon, France}
\author{R. Stoian}
\affiliation{Laboratoire Hubert Curien, UMR CNRS 5516, Université de Lyon,\\ Université Jean-Monnet 42000,
Saint-Etienne, France}
\date{\today}

\begin{abstract}
 The electronic behavior of various solid metals (Al, Ni, Cu, Au, Ti and W) under ultrashort laser irradiation is
investigated by means of density functional theory. Successive stages of extreme nonequilibrium on picosecond timescale
impact the excited material properties in terms of optical coupling and transport characteristics. As these are
generally modelled based on the free electron classical theory, the free electron number is a key parameter. However,
this parameter remains unclearly defined and dependencies on the electronic temperature are not considered. Here, from
first-principles calculations, density of states are obtained with respect to electronic temperatures varying from
$10^{-2}$K to $10^{5}$K within a cold lattice. Based on the concept of localized or delocalized electronic states,  
temperature dependent free electron numbers are evaluated for a series of metals covering a large range of electronic 
configurations. With the increase of the electronic temperature we observe strong adjustments of the electronic 
structures of transition metals. These are related to variations of electronic occupation in localized $d$-bands, via 
change in electronic screening and electron-ion effective potential. The electronic temperature dependence of 
nonequilibrium density of states has consequences on electronic chemical potentials, free electron numbers, electronic 
heat capacities and electronic pressures. Thus electronic thermodynamic properties are computed and discussed, serving 
as a base to derive energetic and transport properties allowing the description of excitation and relaxation phenomena 
caused by rapid laser action.
\end{abstract}


\maketitle \section{Introduction}

 The dynamics of laser-excited materials is an area of intense research as diagnostics of laser-matter experiments can
be discussed by back-tracking the transient properties of the irradiated material. Particularly, the primary phenomena
of transient electronic excitation and energy transport are of utmost importance. Irradiating a metal by a short
laser pulse ($\sim 100$ fs) can lead to a significant rise of the electronic temperature with respect to the ionic
lattice as the energy of the laser pulse can be deposited before the material system starts dissipating
energy by thermal or mechanical ways. The electronic excitation can affect both electronic and structural properties of
the solid, impacting optical coupling, transport and phase transitions. The confinement of the absorbed energy at solid
density pushes the matter into an extreme nonequilibrium state and new thermodynamic regimes are triggered. The
interplay between the ultrafast excitation and the material response still requires a comprehensive theoretical
description for highly excited solid materials including in particular the excitation-dependent band structure evolution
as this influence the response to laser action. \cite{Balling13} Recent advances in studying processes induced by short
laser pulses have revealed the determinant role of primary excitation events. Their accurate comprehension is necessary
to correctly describe ultrafast structural dynamics, \cite{Cavalleri07,Gamaly11} phase transitions,
\cite{Povarnitsyn09,Recoules06} nanostructure formation, \cite{Colombier12} ablation dynamics,
\cite{Lorazo06,Colombier12b} or strong shock propagation. \cite{Zhakhovsky11} In such nonequilibrium conditions,
conduction electrons participating to energy exchange are expected to evolve in time, depending on the excitation
degree. \cite{Kirkwood09} They largely determine the material transient properties and transformation paths. In this
context, they are a crucial parameter required to describe complex ultrafast phenomena involving relaxation of excited
states. Particularly, pure electronic effects (population and band distribution) determining transient coefficients
before structural transitions set in are of interest and we will follow excitation influence in the form of
nonequilibrium electronic temperature.

 At the very beginning of the irradiation process, excited electrons are unhomogeneously distributed within the
electronic band structure of the materials. By collisions and energy transfer, they fastly reach a Fermi-Dirac
distribution. The electronic subsystem is thermalized and the concept of electronic temperature ($T_e$) can be applied.
Mueller \textit{et al.} have recently shown that the electron subsystem thermalizes within a characteristic time $\tau$
in the range of tens of femtoseconds for $T_e$ larger than $10^{4}$K. \cite{Mueller13} Then, a first relaxation channel
is the energy transfer between electronic and vibrational excitations. This is commonly described by a two temperature
model (TTM) \cite{Kaganov57,Anisimov74} based on the assumption that the occupation of electronic and phonons states can
be separately described by two effective temperatures, the electronic $T_{e}$ and the ionic temperature $T_{i}$. In
standard approach, energy transfer between electrons and ions can be modeled by the product of the electron-phonon
coupling parameter $\gamma$ and the temperature difference. This relaxation occurs in the picosecond timescale.
\cite{Sun93} Thus, there exists a period of time where the ionic temperature remains low while electrons fastly reach a
thermalized state of high electronic temperature. In this case, strong alteration of electronic properties preceeds
structural transformation, with consequences on the efficiency of energy deposition. Such nonequilibrium states can be
modelled in the framework of first-principles calculations, by interrogating the electronic influence at various
degrees of electronic heating, while disregarding in a first approximation, the ionic temperature effects.

 Experimentally, excited solids in steady-state cannot be created and thermal nonequilibrium data in these particular
conditions are difficult to be determined from integrated or time-resolved measurements. \cite{Fann92,Wang94,Ao06} In
such complex conditions, simple estimations and models are used to access the behavior of intrinsic material properties.
A strong need then exists to perform multiscale calculations both in space and in time, capable of replicating the
observed behaviors and to predict material response under excitation. Most of the macroscopic models and behavior laws
are based on the picture of free electrons commonly used to describe metals or even dielectric dynamics under laser
irradiation. \cite{Balling13,Gamaly11,Anisimov74,Rethfeld02} \textit{Ab initio} microscopic calculations can supply
macroscopic approaches (optical, thermal, hydrodynamical or mechanical) with implicit dependencies on material 
properties and electronic band structures. Since the density of free electrons $n_{e}$ is not an observable variable in 
a quantum mechanical perspective, its dynamics in thermally excited solids remains poorly explored whereas the transport
parameters and measurable dynamics are commonly depicted and fitted by laws depending on $n_{e}$. 

 It has already been shown in previous works that electronic structures determine the thermodynamic functions and
scattering rates of the heated electron subsystem. \cite{Lin08} This work analyses and extends the largely used
electronic thermodynamic properties derived from the free electron gas model by interrogating the evolution of the free
carriers. This model, based on the assumption of free and non-interacting electrons, works satisfactory in case of
simple metals (Na, Mg, Al...). However it cannot encompass the complexity observed for transition metals, where
$d$-electrons with a higher degree of localization than $sp$-electrons can still participate to optical processes. This 
indicates a potentially important role of electron confinement within more or less diffuse orbitals. Moreover, it has 
already been shown that the increase of the electronic temperature strongly affects the shape of the $d$-band. 
\cite{Recoules06} Under such strong modifications of the electronic systems, it is important to extract from 
calculations an effective free electron number per atom $N_{e}$ classically defining $n_{e}=N_{e} n_{i}$. This effective 
parameter can have importance whenever experimental optical or thermal properties are derived and used to extract other 
parameters such as temperature, stress or conductivity from a nonequilibrium solid. The objective of the investigations 
presented here is to quantify the effects of thermal activation energy $\sim k_{B}T_{e}$ around Fermi energy on $N_{e}$ 
consistently with a rigorously calculated band-structure accounting for Fermi smearing and $d$-band shifting within the 
range of $0.01\leq T_{e}\leq 10^{5}$K.

 We report results from a systematic study on DOS energy broadening performed on a free electron like metal (Al) and on
transition metals (Ni, Cu, Au, W and Ti), some of them with noble character. Section II is devoted to the calculation
and procedure details. In section III, DOS dependence on the electronic temperature is discussed. We focus on the
observable energetic shift and narrowing of the $d$-band and the implications on the chemical potential. Finally, in
order to obtain the electron density relevant to light absorption, heat flux, or mechanical stress induced by electronic
heating, an estimation of the number of free electron per atom, based on delocalized states considerations, is
calculated at all electronic temperatures and discussed for all metals in Section IV. Concluding remarks on the effects
of $n_{e}$ evolution on energetic and transport parameters, especially on electronic pressure and electronic heat
capacity are made in section V.

\section{Calculations details}

 Calculations were done in the framework of the density functional theory (DFT), \cite{DFT1,DFT2} by using the
Abinit package \cite{Abinit} which is based on a plane-waves description of the electronic wave functions. Projector 
augmented-waves atomic data \cite{PAW1,PAW2,PAW3} (PAW) are used to model nucleus and core electrons. The generalized 
gradient approximation (GGA) in the form parameterized by Perdew, Burke and Ernzerhof \cite{GGA-PBE96} or the local 
density approximation (LDA) functional developed by Perdew and Wang \cite{LDA-PW92} are considered for the exchange 
and correlation functional. Semicore electronic states are included in PAW atomic data of Ti and W as they 
significantly improve the description of material properties. The Brillouin zone was meshed with Monkhorst-Pack 
method, \cite{Monkhorst76} with a 30 $\times$ 30 $\times$ 30 $k$-point grid. From the studied metals, only Ni is 
expected to have ferromagnetic properties, but our calculations showed that magnetic properties vanish above $T_e$ = 
$3\times10^3$K, thus, all calculations were done without using spin-polarized methods. To ensure high accuracy of 
calculations, lattice parameters were relaxed up to the point where stress goes beyond $10^{-4}$ eV/\AA, with a cutoff 
energy of 40 Ha.

\begin{table*}[ht]
\caption{Electronic structure of atoms and theoretical, experimental and relative error (\%) of lattice parameters
(\AA) and bulk moduli (GPa) of metal phases at ambiant conditions.}
\label{basic-properties}
\begin{ruledtabular}
\begin{tabular}{llllcccccc}
 Elt  &   XC Functionals       &            Elec. Struc.          &  Chem. Struc. & l$_{th}$ & l$_{exp}$ & Rel. Err.  &
B$_{th}$ &
B$_{exp}$ & Rel. Err.  \\
\hline
 Al   &  GGA           & 3$s^{2}$3$p^{1}$                 &     FCC       &   4.04     &   4.05    &  -0.4     &
  79    &     81    &    -2     \\
 Ni   &  GGA           & 3$d^{8}$4$s^{2}$                 &     FCC       &   3.51     &   3.52    &  -0.3     & 
 192    &    191    &     1     \\
 Cu   &  GGA           & 3$d^{10}$4$s^{1}$                &     FCC       &   3.64     &   3.61    &   0.6     & 
 142    &    133    &     6     \\
 Au   &  LDA           & 5$d^{10}$6$s^{1}$                &     FCC       &   4.05     &   4.08    &  -0.7     &      
 195  &      167    &     14     \\
\multirow{2}*{Ti}  &  \multirow{2}*{GGA}           & \multirow{2}*{3$s^{2}$3$p^{6}$4$s^{2}$3$d^{2}$} & HCP (a,b)    &
  2.93     & 2.95   & -0.6   &   \multirow{2}*{112}  &   \multirow{2}*{114}  &  \multirow{2}*{2}      \\
      &                &                                  & HCP (c)       &   4.66     &   4.69    &  -0.6     &       
  &         &           \\
 W   &  GGA            & 5$s^{2}$5$p^{6}$4$f^{14}$5$d^{4}$6$s^{2}$ &     BCC       &   3.18     &   3.17    &   0.7    
&   295    &    296    &    -5     \\
\end{tabular}
\end{ruledtabular}
\end{table*}

 Al, Ni, Cu, Au, Ti, W cristallize in different phases depending on the environment conditions. Here, we focus on 
cristal structures adopted by metals at ambiant conditions, namely: face-centered cubic (FCC, space group
$Fm\bar{3}m$, 225) structure for Al, Ni, Cu and Au; hexagonal close-packed (HCP, space group $P6_{3}/mmc$, 194) 
structure for Ti; and body-centered cubic (BCC, space group $Im\bar{3}m$, 229) structure in the case of W. Once 
structures are chosen, the accuracy of PAW atomic data is tested through calculations at $T_i = T_e= 0$K conditions 
of lattice parameters and bulk moduli using Birch-Murnaghan equation of states. A good agreement is found between 
our calculated values and experimental data, that are provided in Table \ref{basic-properties}. Some differences are 
noticeable between computed and experimental bulk moduli, \cite{Gaudoin02,Dewaele04,Zhang08} especially when zero-point 
phonon effects are not taken into account. \cite{Gaudoin02} This confirms the reliability of the used PAW atomic data. 
The theoretical lattice parameters computed at this step are then used to calculate $T_e$ dependent DOS. 

 To model laser irradiation effects, we consider timescales where the electrons are thermalized and their distribution
can be described by electronic temperatures. Calculations were done with $T_e$ ranging from $10^{-2}$K to $10^{5}$K 
while $T_i$ remains equal to $0$K. A number of 40 bands per atom is used to ensure a maximum occupation below $10^{-4}$ 
electrons of the highest energy band at $10^{5}$K. $T_e$ dependent DFT calculations are performed following the 
generalization of the Hohenberg and Kohn theorem on many-body systems to the grand canonical ensemble as proposed by 
Mermin. \cite{Mermin65} The finite electronic temperature is taken into account by considering a Fermi-Dirac 
distribution function applied to the Kohn-Sham eigenstates, ensuring a single thermalized state of electrons during the 
self-consistent field cycle. This involves a $T_e$ dependent electronic density and an electronic entropy part  in the 
free energy potential with implicit and explicit dependencies. \cite{Abinit,Khakshouri08} Equilibrium electronic density 
at a finite electronic temperature is determined by minimizing the free energy, the variational functional here, 
resulting in a $T_e$ dependent electronic structure.

\section{$T_e$ effect on Density of States}

\begin{figure}[]
\includegraphics[width=8.4cm]{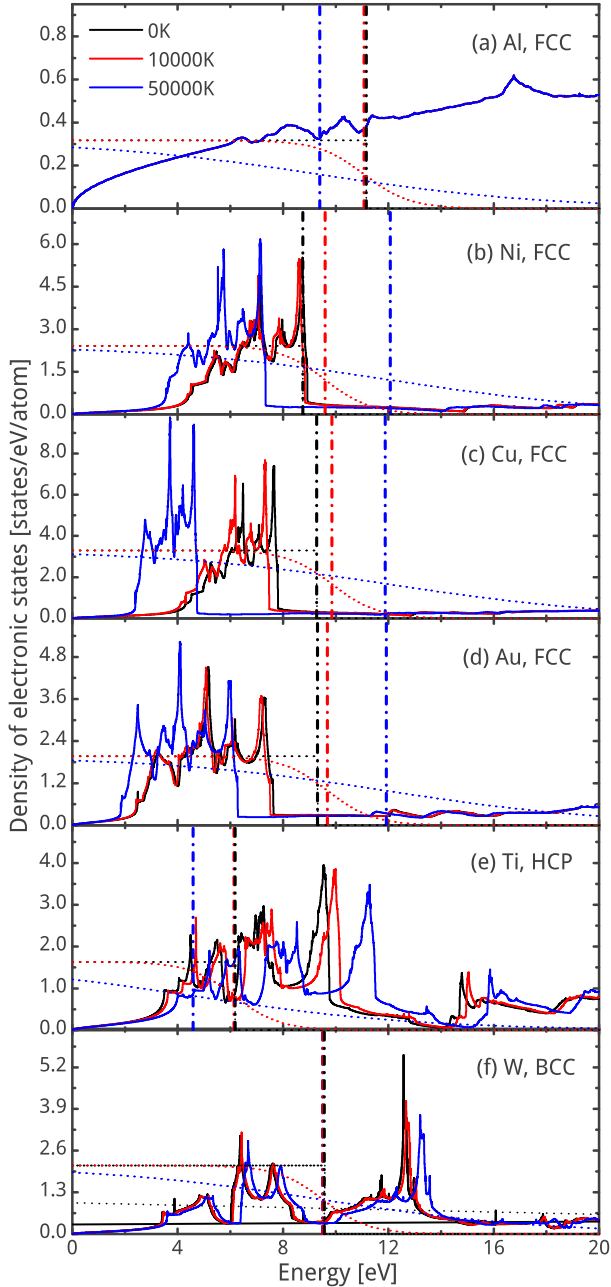}
\caption{\small (color online). Electronic density of states (solid lines), associated Fermi-Dirac distribution
functions (dotted lines) and corresponding electronic chemical potential (dashed lines) for all studied metals. Data
for the following electronic temperatures are shown: $10^{-2}$K (black), $10^{4}$K (red) and $5 \times 10^{4}$K (blue)
curves. \cite{dynopt}}
\label{total-dos}
\end{figure}

 The following discussion is based on the precise determination of the electronic density of states of all the
considered metals. Calculations are performed at twelve different electronic temperatures, from $10^{-2}$ to $10^{5}$K. 
\cite{dynopt} The DOS and associated Fermi-Dirac electronic distribution functions of the discussed metals are shown in 
Fig. \ref{total-dos} at the electronic temperatures of $10^{-2}$, $10^{4}$ and $5\times10^{4}$ K. Here, for simplicity, 
the beginning of the valence band of the DOS was set at 0 eV for each $T_e$. According to this representation, the 
number of valence electrons $N^{v}_{e}$ can be expressed as:

\begin{eqnarray}
   \displaystyle N_{e}^{v} = \int_{0}^{\infty} \! g(\varepsilon,T_{e}) f(\varepsilon,\mu,T_{e}) \,
\mathrm{d}\varepsilon,
\end{eqnarray}

 \noindent where $g(\varepsilon)$ is the DOS and $f(\varepsilon,\mu,T_{e})$ is the Fermi-Dirac distribution
($f(\varepsilon,\mu,T_{e}) = \left\{\exp[(\varepsilon - \mu(T_{e})) / (k_b T_{e})] +1 \right\} ^{-1}$).

 If we first focus the discussion on the DOS obtained at $T_e$=0K, we can notice that they are similar to previous
works. \cite{Papaconstantopoulos86, Recoules06,Lin08} For Al, the DOS adopts the shape of square root function of the
energy, characteristic for a free electron like metal. With transition metals, the $d$-band appears with a typical
$d$-block having a much higher density. This $d$-block is filled or almost filled in case of Ni, Cu and Au, while the
filling is roughly $1/3$ and $1/2$ in the case of Ti and W as showed by the location of the Fermi energy. Generally
speaking, the $d$-bands are narrow in the case of Ni, Cu and Au since almost all $d$-states are filled, which leads to a
weak and non-directional character of the $d$-bonding. \cite{Ponomareva12} At the opposite, they are much more expanded
with the presence of pseudo-gaps in the case of Ti and W, exhibiting a stronger and more directional $d$-bonding.
\cite{Ponomareva12}

\subsection{Shift and shrinking of the $d$-block} 

\begin{figure}[]
\includegraphics[width=8.4cm]{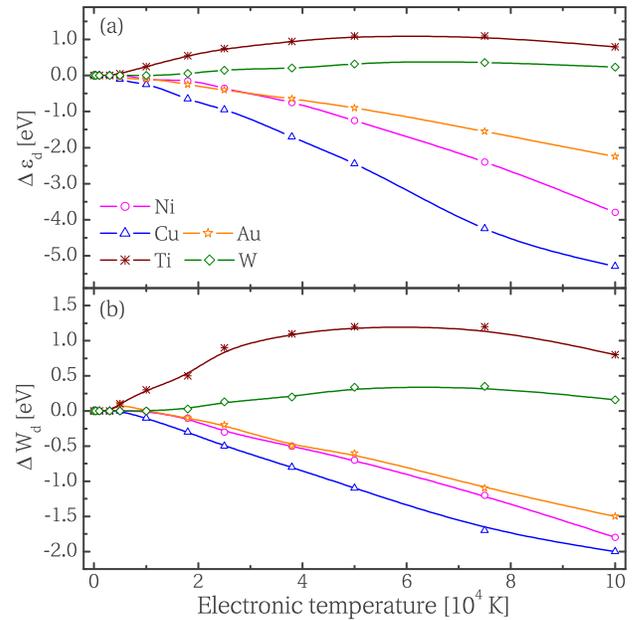}
\caption{\small (color online). Relative changes of the $d$-block center (a) and width (b) with the electronic
temperature, for all studied transition metals.}
\label{d-block-evolution}
\end{figure}

 Density of states exhibit different dynamics when the electronic temperature increases. In the case of Al for example,
the DOS is almost insensitive to $T_e$. This constant behavior of the electronic structure was already noticed in
Ref. [\onlinecite{Recoules06}]. On an other hand, transition metals exhibit more complex DOS due to the
presence of $d$-bands. Metals with $d$-block fully or almost fully occupied by electrons (Ni, Cu and Au) exhibit a
shrinking and a strong shift of the $d$-block toward lower energies when $T_e$ increases. On the contrary, metals with
partially filled $d$-block (Ti and W) display an expansion and a shift toward higher energies of their $d$-block when
$T_e$ is increased. In order to quantify these phenomena, we show in Fig. \ref{d-block-evolution} the relative change of
the $d$-block center [$\Delta \varepsilon_{d}(T_e) = \varepsilon_{d}(T_e) - \varepsilon_{d}(0)$] as well as the relative
change of the $d$-block width [$\Delta W_{d}(T_e) = W_{d}(T_e) - W_{d}(0)$] with $T_e$. For simplicity, they are
deduced by considering a rectangular band model, whose sides are evaluated from the side slope of the electronic density
surrounding the $d$-block. Then, the $d$-block center and width are estimated as $\varepsilon_{d}(T_e) =
(\varepsilon_{d}^{r} + \varepsilon_{d}^{l})/2$ and $W_{d}(T_e) = \varepsilon_{d}^{r} - \varepsilon_{d}^{l}$, where $r$
and $l$ superscripts correspond to the right and left sides of the rectangle.

\begin{figure}[]
\includegraphics[width=8.4cm]{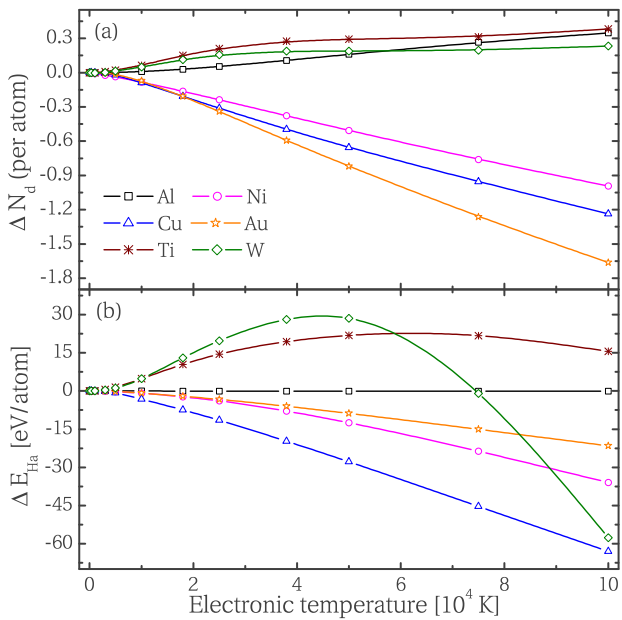}
\caption{\small (color online). Changes of the number of $d$-electrons (a) and changes of the Hartree energy (b) with
respect to the electronic temperature.}
\label{HandDchanges}
\end{figure}

 The $d$-block modifications observed here for Au were already reported in Ref. [\onlinecite{Recoules06}], with an
explanation of the phenomenon based on changes of the electronic screening. When an electronic temperature is applied,
the depopulation of 5$d$-block leads to a decrease of the electronic screening which makes the effective electron-ion
potential more attractive. \cite{Recoules06} The consequence is a global shift of the electronic states toward lower
energies. A similar effect likely occurs in the case of Ni and Cu even if the depopulation is now concerning
$3d$-electrons instead of $5d$ as in Au. Considering the shift toward higher energies and extension of the $d$-block in
case of Ti and W, one should expect an increase of the screening effect with the augmentation of $T_e$ for these two
metals. To validate this assumption, the change of the number of $d$-electrons $\Delta N_d$ and the changes of Hartree
energies $\Delta E_{Ha}$ as a marker of the changes of electronic localization have been evaluated and are presented in 
Fig.\ref{HandDchanges}. We specify that the concept of electronic localization refers to a certain degree of spatial 
concentration of the charge density. The relevance of these observations are discussed below.

 Except the case of Al, all considered metals have electronic configurations in the form of (n-1)$d^x$n$s^y$. Since
the main quantum number is one of the dominant parameter characterizing the diffuse nature of an orbital, 
\cite{Waber65} (n-1)$d$-orbitals are more localized and overlap less than n$sp$-ones. Consequently, the resulting 
$d$-band is confined energetically with higher density of state. Electrons occupying this band are also experiencing 
spatial localization. Thus, one can expect that changes in the electronic screening mainly come from changes of the 
electronic occupation of this $d$-band. To gather deeper insight, we computed the number of $d$-electrons from 
integrations of angular-momentum projected DOS, and we plotted $\Delta N_d$ as a function of $T_e$ in Fig. 
\ref{HandDchanges}a. As expected, $\Delta N_d$ decreases for Ni, Cu and Au. $T_e$ depopulates the $d$-band, which leads 
to a decrease of the electronic screening as discussed previously. At the opposite, $\Delta N_d$ increases for Ti and W. 
This is a consequence of partially filled $d$-block, since electronic excitation depopulates both $sp$-bands and the 
bottom part of the $d$-band and populates the top part of the $d$-band. Semicore electronic states also undergo an 
electronic depopulation at $T_e$ above $5\times10^{4}$K. For example, at $10^{5}$K, the depopulation reaches 0.3 
electrons for 3$p$-semicore electronic states of Ti and 0.4 electrons for 4$f$-semicore electronic states of W. 
Consequently, the total number of $d$-electrons increases in agreement with a strengthening of the electronic screening 
at least up to $5\times10^{4}$K, where the depopulation of semicore states seems to moderate the effect, as we can see 
on Fig. \ref{d-block-evolution} with a decrease of shifts and width changes. Finally, the behavior of Al is particular 
with an increase of $N_d$ and no effects on the density of states. This is related to the occupation of a high energy 
$d$-band that does not contribute to the electronic localization.

 However, DOS angular-momentum projection methods suffer from some drawbacks. Firstly, they are performed on spheres
centered on atoms and there is always some intersphere informations lost during the process. Consequently, some
electronic states and some electrons are not accounted for. Secondly, projections on angular-momentum do not allow
distinctions between 3$d$, 4$d$ or 5$d$ bands. Then, the number of $d$-electrons computed can be the summation of
electrons from localized (n-1)$d$ band with n$d$ or even (n+1)$d$ delocalized bands. Thus, the variation of
$d$-electron numbers has some uncertainties, but still provides information about trends. In order to get a more precise
perspective of how electronic screening is affected by $\Delta N_d$, we rely on Hartree energy. The changes of Hartree
energies with respect to $T_e$ are plotted in Fig. \ref{HandDchanges}b. We recall that this energetic term is the
Coulomb repulsive self-energy of the electronic density: $E_{Ha}= \frac{1}{2} \iint \! \frac{n(r)n(r')} {|r-r'|} \,
\mathrm{d}r \mathrm{d}r'$. It corresponds to the global electron-electron interaction. In a perfectly homogeneous
spatial distribution, this quantity would reach a minimum value, however, constrained by the electronic structure,
electrons are not homogeneously distributed. Specifically, semicore states and (n-1)$d$-bands strongly concentrate 
the charge density, which generates an electronic screening of the nucleus. Variations of the electronic occupation of 
these electronic states with $T_e$ impact Hartree energies, which can be related to a change of electronic screening.
Thus, the evolution of $\Delta E_{Ha}$ with $T_e$ is an indicator for the gain or the loss of electronic localization.
This will be mainly the consequence of $N_d$ changes due to the spatial confinement of the (n-1)$d$-band we discussed 
above. The evolution of $\Delta E_{Ha}$ also reflects the change of electronic screening of the ions. On Fig.
\ref{HandDchanges}, we can notice a good correlation between $\Delta N_d$ and $\Delta E_{Ha}$ with $T_e$ at least up to
temperatures of $5\times10^{4}$K. When the number of $d$-electrons decreases there is a loss of electronic localization.
This leads to a decrease of $E_{Ha}$ which signals a decrease of electronic screening. The opposite phenomenon occurs
when the $N_d$ increases, with a gain of electronic localization leading to an increase of $E_{Ha}$ correlated to an
increase of electronic screening. Finally, $\Delta E_{Ha}$ of Al remains roughly equal to zero, while the electronic
occupation of high energy bands increases. It confirms that high energy bands, including $d$-bands, are sufficiently
delocalized to be considered as uneffective on electronic localization.

 From $E_{Ha}$ and $N_d$ changes, some unexpected behaviors are also observed. Firstly, $\Delta N_d$ is correlated to
$\Delta E_{Ha}$ in the case of Cu and Ni. The stronger is the decrease of $\Delta N_d$, the stronger is the decrease of
$\Delta E_{Ha}$. However, this rule is no longer available when considering Au. This is due to the fact that
$d$-electrons belong to the 5$d$-orbitals in the case of Au whereas they belong to 3$d$-orbitals in the case of Cu and
Ni. Associated to more diffuse orbitals, $d$-electrons of Au are already less localized than those of Cu and Ni, and it
leads to a lower loss of localization when $d$-band is depopulated with the increase of $T_e$. In other words, $\Delta
E_{Ha}$ is lowered in the specific case of Au since its $d$-electrons are already less localized. Secondly, $\Delta
E_{Ha}$ is not correlated to $\Delta N_d$ in case of Ti and W at electronic temperature higher than $5\times10^{4}$K.
Whereas the number of $d$-electrons still increases, the Hartree energy decreases, with a particularly strong decrease
for W. This is attributed to the electronic depopulations of semicore electronic states that occur at high $T_e$
as discussed above. The strong evolution of Hartree energy for W originates from the significant change of electronic
screening generated by the depopulation of highly localized 4$f$-electrons. We assume here that semicore electrons are
fully thermalized with valence electrons even if semicore thermalization timescales are difficult to estimate. Being low
in energies, these semicore states are not directly excited by laser irradiation and thermalization time is dependent on
electron-electron collision frequency. \cite{Mueller13,Fisher05} At $10^{5}$K, Fisher \textit{et al.} \cite{Fisher05}
estimate about 1-10 sucessfull impact probabilities during irradiation timescale, for binding energies of semicore
electronic states similar to the ones considered here. This suggests a short thermalization time but accurate
description is beyond the scope of the present work. 

\begin{figure}[]
\includegraphics[width=8.4cm]{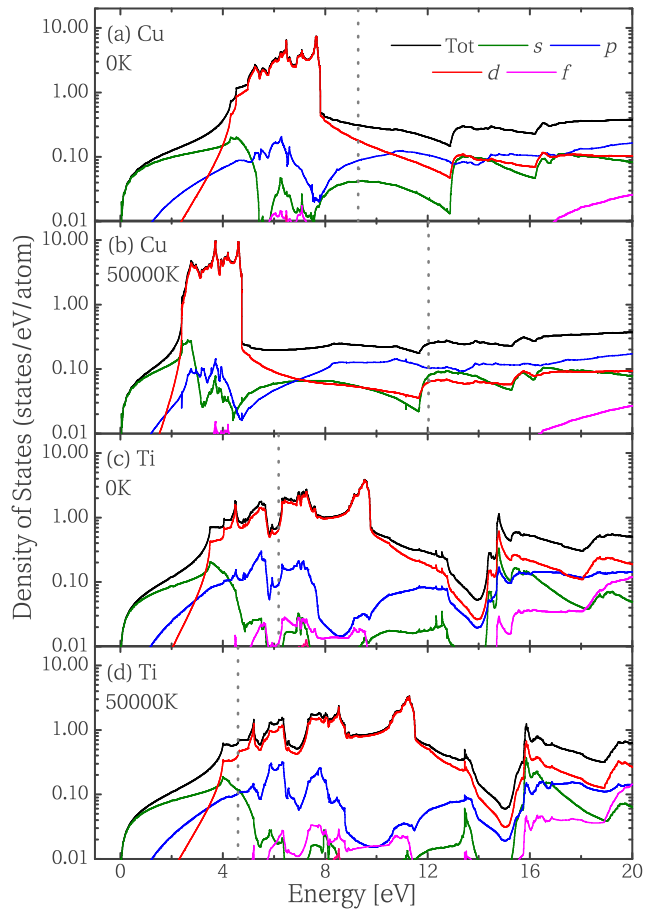}
\caption{\small (color online). Evolution of the total DOS and $spdf$-components of the DOS for Cu at $0$K (a) and $5
\times 10^{4}$K (b), and for Ti at $0$K (c) and $5 \times 10^{4}$K (d). Dashed lines correspond to Fermi levels.}
\label{Cu-Ti-pdos}
\end{figure}

 Generally speaking, we note a relatively good agreement between $\Delta N_d$ and $\Delta E_{Ha}$ on one side, and 
$\Delta \varepsilon_{d}$ and $\Delta W_{d}$ on the other side. For Ni, Cu and Au, the decrease of $\Delta N_d$ leads to 
a loss of electronic localization that generates a decrease of $\Delta E_{Ha}$. The effect can be related to the 
decrease of the electronic screening and thus to the increase of the global electron-ion effective potential that 
shifts electronics states of these metals toward lower energies. This shift applies non-homogeneously on the $d$-block 
since bottom $d$-block is less affected by depopulation and thus by changes of electronic screening than the top part, 
as can be seen on Fermi-Dirac distributions on Fig. \ref{total-dos}(b-d). As a result the $d$-block is shrinked, and 
$\Delta \varepsilon_{d}$ and $\Delta W_{d}$ decrease. Shifts also apply to electronic states of higher energies, as 
shown for Cu on Fig. \ref{Cu-Ti-pdos}(a,b). For Ti and W, the increase of $\Delta N_d$ leads to an increase of $\Delta 
E_{Ha}$. The corresponding gain in electronic localization produces a stronger electronic screening. As a result, the 
electron-ion effective potential is less attractive and bands are shifted toward higher energies. It also applies 
inhomogeneously to the $d$-block with its bottom states less affected by changes of electronic population than its top 
states, as can be seen on Fermi-Dirac distributions on Fig. \ref{total-dos}(e,f). As a consequence, the $d$-block 
extends and is shifted toward higher energies, and $\Delta \varepsilon_{d}$ and $\Delta W_{d}$ increase. Higher energy 
electronic states are also affected by this increase of electronic screening, as we can note with the shift of other 
bands toward higher energies for Ti on Fig. \ref{Cu-Ti-pdos}(c,d). This discussion synthesizes $T_e$ effects on DOS for 
a range of representative metals, with various possible impacts on electronic properties.

\subsection{Electronic distributions}

\begin{figure}[]
\includegraphics[width=8.5cm]{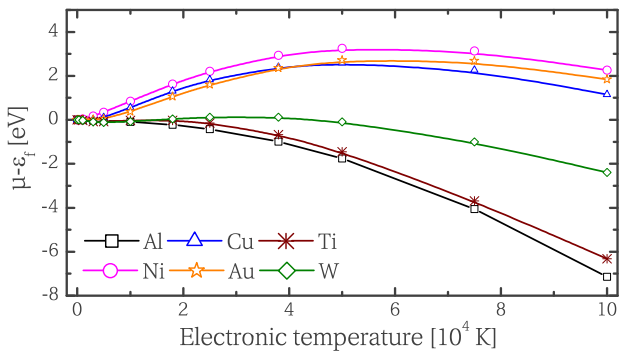}
\caption{\small (color online). Electronic chemical potential evolution with the electronic temperature for
all studied metals.}
\label{electronic-chem-pot}
\end{figure}

 As discussed above, the electronic distribution is ensured by the Fermi-Dirac function characterized by the electronic
chemical potential $\mu(T_e)$. As already showed by Lin \textit{et al.}, \cite{Lin08} the electronic chemical potential
exhibits strong variations, depending on the material studied. More precisely, these variations are related to the
assymmetric distribution of the density of electronic states from both sides of the Fermi energy. For this reason, the 
electronic chemical potential moves toward higher energies in the case of Ni, Cu and Au, while it is displaced toward 
lower energies in the case of Al, Ti and W. In the present calculations, the  $T_e$ dependence of the DOS produces 
important shifts of the $d$-block. Since the $d$-block concentrates electronic states, $\mu(T_e)$ is also strongly 
affected by these shifts. In Fig. \ref{electronic-chem-pot}, the relative changes of the electronic chemical potential 
is shown for all studied materials. 

 One important observation has to be made here. Previous works \cite{Lin08} reported evolutions of $T_e$ dependent
properties computed from DOS performed at $T_e = 0$K. In the present case, this approach takes into consideration the
relaxion of DOS with $T_e$, and changes in the electronic structure are impacting the electronic chemical potential. As
a general note, one can observe that trends are similar between the temperature evolution of $\mu$ derived from $T_e$
dependent DOS and $\mu$ originating from $T_e = 0$K DOS. The agreement is very good in the case of Al, that we can
attribute to weak changes of its electronic structure with $T_e$. However, in case of Ni, Cu and Au, the increase of
$\mu$ is lowered in $T_e$ dependent DOS situation, a difference that is due to the shift of the $d$-block with $T_e$.
Since the $d$-block is the main electron reservoir, when it is shifted toward lower energies, the electronic chemical
potential accounts for this evolution and its displacement tends to follow. In the case of Cu for example, at 
$5\times10^4$K, $\mu$ from $T_e$ dependent DOS is decreased by 2 eV compared to $\mu$ from $T_e = 0$K DOS. For W and 
Ti, the  decrease of $\mu$ is also lowered in $T_e$ dependent DOS. This is attributed to the shift of the $d$-block 
toward higher energies, which is accommodated by $\mu$. This indicates the importance of considering $T_e$ effects on 
band structure while regarding the evolution of electronic populations, with direct consequences on the determination 
of free electron numbers.

\section{Free electron numbers}

 The number of free electrons is a quantity difficult to define since the quality of being ``free'' is elusive. In
metals, electrons are implicitly considered as free if they belong to orbitals having the highest main quantum number in
the atomic electronic configuration (EC).\cite{Ashcroft} For example, EC of Al and Cu are respectively 3$s^{2}$3$p^{1}$
and 3$d^{10}$4$s^{1}$, and their corresponding number of free electrons are 3 and 1, respectively. The highest main
quantum number is chosen as it characterizes the diffuse and overlapping nature of the orbitals. Hence, electrons
belonging to those orbitals are assumed to be free of moving in a large space and by extension in the whole metal, with
trajectories limited by collisions and with parabolic dispersion laws. However, the use of atomic electronic
configurations induces a degree of incertitude when applied to condensed phases. In addition, $N_e$ determined from EC
does not allow any change while $T_e$ increases. As a consequence, and motivated by the necessity of giving a certain
evaluation of the free electron quantity, we computed the number of free electrons directly from the electronic
structures of metals.

 In order to improve the determination of $N_e$, we have to distinguish electrons belonging to localized states (assumed
to be non-free) from those belonging to delocalized states (considered as free). For this, it is important to determine
which are the localized states in the density of states. $d$-orbitals from electronic configuration of transition metals
defined by (n-1)$d^x$n$s^y$ are less diffuse than $sp$-orbitals due to lower main quatum number. Consequently, they
overlap less and thus interact less than $sp$-orbitals. The resulting $d$-bands produce a characteristic $d$-block 
of very high density of electronic states. At the opposite, $sp$-bands are delocalized and generate $sp$-bands of low
density with a square root distribution, similar to Al (see Fig. \ref{total-dos}). The difference of density between
localized $d$-block and delocalized $sp$-bands is large and it is then easy to distinguish them in the DOS. Using a
method to remove localized states from the DOS, one can compute free electron numbers according to the previous
description, i.e. electrons occupying delocalized electronic states only. Accordingly, the number of free electrons is 
given by the integration of the DOS weighted by the Fermi-Dirac distribution:

\begin{eqnarray}
   \displaystyle N_{e}=\int_{0}^{\infty} \! g_{deloc}(\varepsilon,T_{e}) f(\varepsilon,\mu,T_{e}) \,
\mathrm{d}\varepsilon,
\end{eqnarray}

\noindent with $g_{deloc}$ being the delocalized part of the density of states only.

\begin{figure}[]
\includegraphics[width=8.5cm]{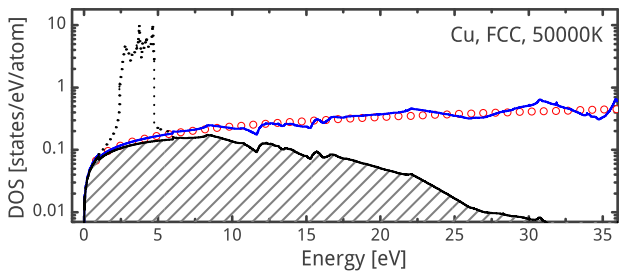}
\caption{\small (color online). The blue curve is the density of delocalized electronic states for Cu at 
$5\times 10^{4}$K. The empty red dots show the square root fit of the DOS and the dotted black curve represents the 
removed part of the initial DOS, constituted of the non-contributing localized $d$-block. The hatched part refers 
to the free electrons and is computed from the integration over the Fermi-Dirac distribution corresponding to Eq. (2).}
\label{Cu-mdos}
\end{figure}

 To remove localized states from the DOS, several methods can be used. One of the most simple implies to fit the density
of states to a curve having a square root shape, as this will artificially remove the high density of states of the
$d$-block. \cite{Loboda11} The square root shape is chosen since it is the distribution of the density of electronic
states adopted by a free electron like metal, as observed for Al on Fig. \ref{total-dos}a. Considering the electronic
structure of $d$-band metals, square root shape is thus a criterium for indentifying delocalized states. Here, in order
to keep all DOS subtleties, this square root fit is only used to replace the $d$-block, resulting in a DOS of 
delocalized states, that can be written as:

\begin{eqnarray}
  \displaystyle g_{deloc}(\varepsilon) = g(\varepsilon) - [g(\varepsilon) - \alpha \sqrt{\varepsilon}]_{dblock},
\end{eqnarray}

\noindent where $\alpha \sqrt{\varepsilon}$ is the fit of the DOS. The correction is only applied to the energy range 
containing the $d$-block. Fig. \ref{Cu-mdos} exemplifies the whole process in the case of the DOS of Cu, indicating the 
fit results at $5\times 10^{4}$K. By proceeding that way, a small part of $d$-electrons also contributes to the free 
electron number. We assume that a small part of the $d$-band can be considered as delocalized too. The analysis of 
Hartree energies (Fig. \ref{HandDchanges}b) already showed that some $d$-electrons are less localized than others (case 
of Au 5$d$-electrons versus case of 3$d$-electrons of Ni and Cu). Moreover, the $d$-band is not restricted to the very 
localized $d$-block, as shown in Fig. \ref{Cu-Ti-pdos}. Part of it easily matches a square root distribution of 
electronic states, indicating non-negligible overlaps and interactions between $d$-components, in a similar way to 
what is observed with $sp$-components. Finally, as previously discussed, $d$-bonds are mainly weak and non-directional, 
\cite{Ponomareva12} thus it would not be surprizing that, despite a strong localization character, part of $d$-electrons 
have an ability to be mobile.

\begin{figure}[]
\includegraphics[width=8.5cm]{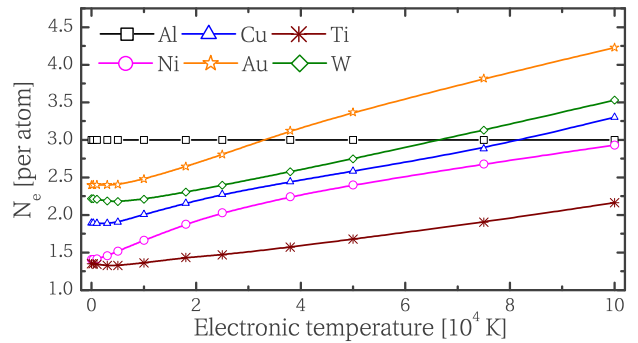}
\caption{\small (color online). Evolution of the number of free electrons per atom ($N_e$) with the increase of $T_e$,
for the considered metals.}
\label{free-electrons}
\end{figure}

 Once they are only made of delocalized states, the DOS can now be integrated and $N_e$ can be deduced. The number of
free electrons per atom for all considered metals is presented in Fig. \ref{free-electrons}. It shows strong variations
as $T_e$ increases, exept in the case of Al, where this number remains constant with $N_e= 3.0$ free electrons per atom.
This is an expected free electron behavior since excited electrons are leaving delocalized states to reach other 
delocalized states. For Ni, Cu and Au, the localized $d$-block can be considered as a reservoir of non-free electrons 
susceptible to be depopulated with $T_e$, depending on the relative location of the Fermi energy. Then, non-free 
electrons from the localized $d$-block reach delocalized states and become free, which leads to an increase of $N_e$ 
with $T_e$. In the case of Ti and W, the partially occupied $d$-block plays an ambivalent role. At low $T_e$ the bottom 
part of the $d$-block is filled of non-free electrons while the top part is empty but consists of localized states that 
can potentially trap excited electrons. As a consequence, $N_e$ remains constant or slightly decreases at low electronic 
temperatures. However, at temperatures above $10^{4}$K, $N_e$ increases as in the case of Ni, Cu and Au, by populating 
delocalized states of higher energy.

 The typical values of $N_e$ deduced only from electronic configurations of isolated atoms are independent on the
electronic temperature. They are provided in Table \ref{free-electrons-0k} alongside with the $N_e$ obtained by the
present approach at 0K. As already mentioned, we obtained a good agreement in case of Al since the electronic structure
is only made of delocalized states and thus $N_e$ remains constant and equal to the number of electrons from EC.
However, differences can be large with respect to other metals, especially since electronic structure of condensed
phases allows the transfer of electrons between bands. They also come from the fact that we considered the $d$-band as
partially delocalized. We observed that the free electron numbers calculated here approach values from other
estimations. \cite{Ebeling91,Kirkwood09} In case of Au, the present $N_e$ is higher than the value typically used. This
is attributed to different Fermi energies, almost twice higher in our case (10.2 eV) than the value generally considered
in the litterature (5.5 eV). \cite{Ashcroft} It should also be noted that prior hypothesis concerning the charge density
led to compensating corrections on other electron parameters related to classical models (Drude), via e.g. the effective
mass used for Au. \cite{Inogamov09}

\begin{table}[]
\caption{Number of free electrons, from Ref. [\onlinecite{Ashcroft}] and from this work obtained at $T_e$ = 0K.}
\label{free-electrons-0k}
\begin{ruledtabular}
\begin{tabular}{lcccccc}
 $N_e$                          &   Al     &   Ni   &   Cu   &   Au   &   Ti   &   W   \\
 Ref. [\onlinecite{Ashcroft}]    &   3      &   2    &   1    &   1    &   2    &   2   \\
 This work                      &   3.0    &   1.4  &   1.9  &   2.4  &   1.4  &   2.2 \\
\end{tabular}
\end{ruledtabular}
\end{table}

\section{Energetic and transport parameters}

 In order to investigate to what level the band structure dependence on the excitation degree (via $T_e$) affects 
macroscopic transient characteristics and physical quantities, we evaluated thermodynamic properties of electrons when
$T_{e}$ differs from $T_{i}$. Such excitation character corresponds to ultrashort pulse laser irradiation of metal free
surfaces, where the spatio-temporal evolution of the electron distribution and its return to equilibrium is calculated
by a Boltzmann formalism \cite{Rethfeld02} or by the two temperature model (electronic thermalization assumed). We
recall that the TTM describes the energy evolution of electrons and ions subsytems using two diffusion equations coupled
by an electron-phonon transfer term.\cite{Anisimov74} The electronic thermal energy gain is connected to the temperature
by the electron specific heat $C_{e}$. An accurate evaluation of this property is  crucial in laser-matter interaction
simulations since it provides a correct estimate of the rise of temperature in the  electron system. TTM enables to take
into account nonequilibrium effects on the kinetics of the material when it is included in classical molecular
dynamics simulations or in two temperatures hydrodynamics approaches. \cite{Lin08b,Colombier12,Inogamov12} The
electronic contribution to $P_{e}$ results directly from the electron heating, depending thus on $T_{e}$ but also on the
density of states. Consequently, swift matter dynamics start especially due to the pressure gradient generated in
the electron system. The key parameters to correctly reproduce the ultrafast nonequilibrium evolution of the material
are then based on an electronic equation of states, connecting electronic specific heat $C_{e}$ and pressure $P_{e}$
with free electron density $n_{e}$ and temperature. $C_{e}$ and  $P_{e}$ are generally derived from the thermodynamic
properties of an ideal Fermi gas. \cite{More88,Eidmann00,Jiang05,Colombier05} This work allows to insert the subtle
effects of DOS modifications allowing more accurate perspectives. Transport properties, i.e. electronic thermal and
electrical conductivity and electron-phonon coupling strength are also important to complete the kinetic equations but
they are beyond the scope of this paper and will be only briefly mentioned. We will focus here on the influence of band
structure evaluation on electronic thermodynamic properties. 

\begin{figure}[]
\includegraphics[width=8.5cm]{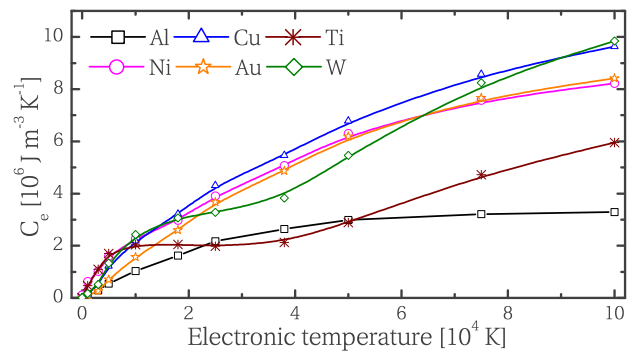}
\caption{\small (color online). Evolution of the electronic heat capacity with respect to the electronic temperature 
for the studied metals. }
\label{electron-heat-capacity}
\end{figure}

 The electron specific heat of metals can be derived with respect to the electronic temperature by $C_{e} = \partial
E/\partial T_{e}$, where $E$ is the internal energy of the electron system. The evolution of the specific heat
under electronic excitation is shown in Fig. \ref{electron-heat-capacity} where several typical behaviors are observed.
For Al, $C_e$ rapidly saturates to the lowest value of all other considered metals. The rest of metals remains far from
saturation and reaches much higher values than Al. For W and Ti, $C_e$ exhibits a first leveling from $10^{4}$K to
$4\times 10^{4}$K and restarts to increase at higher temperatures. At low and intermediate $T_e$ a relatively good
agreement is found with those obtained from $g(\varepsilon)$ evaluated at 0K, where $[\partial g(\varepsilon)/\partial
T_{e}]_{V}$ is neglected. \cite{Lin08} For transition metals, increasing discrepancies appear at higher temperatures,
mainly due to shifts of the $d$-block that are highly affected by $T_e$ increase. As expected, the temperature
dependence of $C_e$ is linear at low electronic temperature and tends to saturate for high $T_e$ toward the
non-degenerate limit $3/2 n_{el}k_b$, where $n_{el}$ includes both free electrons and part of $d$-electrons. As already
mentioned by Lin \emph{et al.},\cite{Lin08} thermal excitation from the $d$-band results in a positive deviation of
$C_e$ from the linear temperature dependence. Similar results are obtained with our temperature dependent calculations,
with an additional deviation resulting from excitation of semicore electrons in case of Ti and W and high
$T_e$.

\begin{figure}[]
\includegraphics[width=8.5cm]{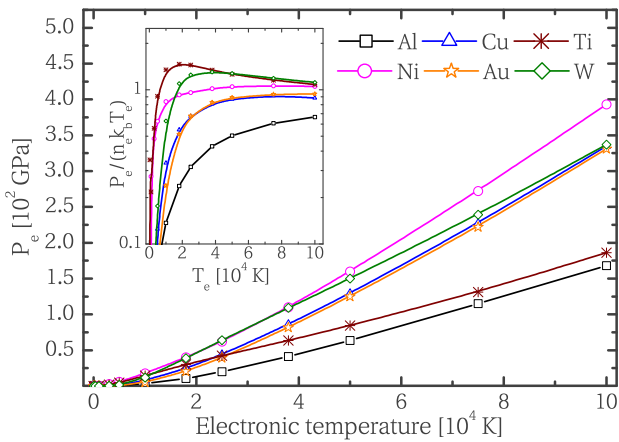}
\caption{\small (color online). Evolution of the electronic pressure with the electronic temperature for all the studied
elements. Inset: renormalization of $P_e$ with respect to the free electron gas pressure limit $n_ek_bT_e$.}
\label{electronic-pressure}
\end{figure}

 The electronic pressure $P_e$ is determined by the derivative of the electronic free energy $F$ with respect to volume 
as $P_e= -\partial F/\partial V= -\partial E/\partial V + T_{e}\partial S/\partial V$, where $S$ is the entropy of the
system. The last term corresponding to the entropy contribution to the pressure has been shown to be largely dominant in
this range of $T_{e}. $\cite{Khakshouri08} $P_e$ evolution for the different metals is plotted in Fig.
\ref{electronic-pressure} and shows that $P_{e}$ increases rapidly as $T_{e}^{2}$ for low excitation then scales as
$T_e$ for higher temperatures. At $2.5\times 10^{4}$K, the electronic pressure is in the order of tens of GPa,
and exceeds 100 GPa at $5\times10^{4}$K for all metals except Al. Finally, at $10^{5}$K more than 300 GPa are reached in
case of Ni, Cu, Au and W while it approaches a level of 200 GPa in case of Al and Ti. The fast increase of the
electronic pressure is of interest since it is likely impacting the stability or the properties of materials and their
evolution upon excitation, notably the initial steps of the thermodynamic trajectories. This strong increase of $P_e$
with $T_{e}$ comes from the occupation of the high energy states by the electrons, which is governed by the electronic
structure and by entropic changes with the electronic temperature.

 In order to exhibit band structure effects and to test our free electron approach, we renormalized $P_e$ with respect
to the free electron gas pressure limit $n_ek_bT_e$. The ratio is plotted in the inset of Fig. 
\ref{electronic-pressure}. At low electronic temperatures, degeneracy and band structure effects are dominating and the 
curves are far from the value of unity, that would characterize an ideal non-degenerated electron gas behavior. 
However, at higher electronic temperatures, curves tend to saturate at the value of 1.0, which indicates that the free 
electron numbers we have derived are consistently characterizing the electronic pressure of the system. This asymptotic 
behavior is not achieved using constant values of $n_e$ given by Ref. [\onlinecite{Ashcroft}]. On an other hand, the 
effect of the entropy contribution on the renormalized $P_e$ is enhanced. The entropy contribution reflects the 
electronic disorder centered around the Fermi level. It comes from a compromise between the number of available 
electronic states and the number of electrons allowed to fill these states. In this context, when the Fermi level is 
within the $d$-block, the entropy effect is stronger, as for Ti and W, than when it is located somewhere else in the 
DOS (as in the case of Ni, Cu, Au). Al shows the lowest renormalized values due to lowest density of states.

 Based on these results, an equation of states of the electron system can be constructed, including the interactions
with a static array of positive ions but neglecting interactions with phonons. To predict laser light absorption,
subsequent material heating and eventual phase transitions, the transport model should take into account the observed
changes in band structure. In a first approach, transport properties can be estimated based on classical formulas but
including the evolution of $n_{e}$ with $T_{e}$. In this way, $k_{e}$ conductivity can be estimated roughly based on 
the relation derived from the kinetic equation between the electron heat capacity and $k_{e}$. \cite{Petrov13} If 
$\tau_{e}$ is the relaxation time between collisions and $v_{e}$ the average electron velocity, then the mean free path 
$l_{e}=v_{e} \tau_{e}$ allows to make the standard connection which writes $k_{e}(T_{e})=\frac{1}{3} C_{e}(T_{e}) l_{e} 
v_{e}$. Concerning the laser absorption, insights from the classical formalism of an electron in an optical electric 
field, i.e. the Drude-Lorentz model, can also be obtained from the calculated properties. Whereas the energy levels and 
their own occupation define the response of the material to applied optical fields, it is possible to describe the 
dielectric function by intraband and interband contributions. The optical properties are related to the structure and 
the electronic configuration of the material and the interband is highly sensitive to the DOS derived in section III. 
On the other hand, the intraband part (Drude) only depends on the number of free carriers per atom and on an effective 
momentum scattering time $\tau$ as $\sigma=n_{e}e^{2}\tau/m(1-i\omega \tau)$, where $\omega$ is the laser angular 
frequency. In this way, it becomes possible to make reasonable deductions about excitation when optical property changes 
are measured. \cite{Colombier08} These insights could become more accurate if we consider informations based on 
electronic structure calculations, with higher computational costs in this case. \cite{Holst13} Simpler models based on 
$n_{e}$ and interband transition from filled to empty states could be useful as a complementary approach, with 
nevertheless a lower accuracy.

\section{Conclusion}

 In the present study, $T_e$ dependent density functional calculations were performed on a representative range of
metals: Al, Ni, Cu, Au, Ti and W, with simple and transition character. Electronic temperatures from $10^{-2}$ to
$10^{5}$K were used to evaluate electronic properties in nonequilibrium conditions.

 In a first step DOS modifications with $T_e$ are discussed. It is shown that almost all bands are affected by energy
shifts, but the most affected states are those involving the localized part of the $d$-band which characterizes the
transition metals. Shifts towards lower energies and shrinking are observed for filled or almost filled $d$-block
metals, illustrated by Ni, Cu and Au. Shifts towards higher energies and extensions for partially filled $d$-block, with
Ti and W as examples. All these modifications are explained by evolutions of electron-ion effective potential that
result from variation of the electronic screening generated by changes of the electronic occupation of the localized
$d$-block. This was validated by two consistent approaches, the evolution of the number of $d$-electrons and the
modification of the Hartree energies with $T_e$.

 Changes of the DOS with electronic temperatures impact electronic properties, like the electronic chemical potential
and the electronic heat capacities that are discussed and compared to previous calculations performed at $T_e = 0$K. A
good agreement is obtained at low and intermediate temperatures, while an increasing discrepancy is observed when shifts
within the electronic structures become stronger at high temperatures. The concept of electronic pressures is also
addressed, with pressures rapidly reaching high values, of tens or hundreds of GPa, questioning material stability as
$T_e$ increases.

 Free electron numbers, dependent on the electronic temperature, are also computed from DOS. They are defined as
belonging to delocalized states characterized by density of states having a square root energy dependence. As
expected, for a free electron like metal as Al, this number remains constant. However, $N_e$ always increases for
transition metals, with specific behaviors depending of the filling degree of the $d$-block. Ni, Cu and Au exhibit a
$N_e$ that rapidly increases with $T_e$ while a small lag is observed for Ti and W, with an increase at higher 
temperatures. At high temperatures, these free electron numbers are found consistent with $P_e$ in an ideal gas limit.

 Finally transport properties were addressed via the evolution of the electronic heat capacity and the electronic
pressure with $T_e$. The electronic heat capacity links the quantity of absorbed energy to the electronic temperature,
giving an access to the number of free electrons. On the other hand, the rapidly growing electronic pressure may impact
phase stabilities. Both are crucial properties, since significant deviations of the thermophysical properties of metals
from the commonly used approximations may have important practical implications in ultrashort laser material processing
applications.

\section{Aknowledgments}

 We acknowledge Marc Torrent for providing efficient PAW atomic data. This work was supported by the ANR project
DYLIPSS (ANR-12-IS04-0002-01) and by the LABEX MANUTECH-SISE (ANR-10-LABX-0075) of the Université de Lyon, within the
program "Investissements d'Avenir" (ANR-11-IDEX-0007) operated by the French National Research Agency (ANR). Part of the
numerical calculations has been performed using resources from GENCI, project gen7041.

\end{document}